\documentclass[aps,pra,floatfix,showpacs,noshowkeys,epsfig,graphics]{revtex4}%
\usepackage{graphicx}
\usepackage{amsmath}
\usepackage{amsfonts}
\usepackage{amssymb}

\newcommand{\newc}{\newcommand}
\newc{\beq}    {\begin{equation}}
\newc{\eeq}    {\end{equation}}

\newc{\beqa}    {\begin{eqnarray}}
\newc{\eeqa}    {\end{eqnarray}}
\newc{\bs}    {\section}
\newc{\no}    {\\ \nonumber}

\newc{\st}    {\stackrel}

\begin{document}
\title{  Are galaxies extending?}
\author{Jae-Weon Lee}\email{scikid@jwu.ac.kr}
\affiliation{School of Computational Sciences,
             Korea Institute for Advanced Study,
             207-43 Cheongnyangni 2-dong, Dongdaemun-gu, Seoul 130-012, Korea}
 \affiliation{ Department of energy resources development,
Jungwon
 University,  5 dongburi, Goesan-eup, Goesan-gun Chungbuk Korea
367-805}
\date{\today}
\begin{abstract}
  It is suggested that the recently observed size evolution of very massive compact  galaxies
in the early universe can be  explained, if dark matter is in Bose Einstein condensate.
   In this model the size of the dark matter halos and galaxies  depends
   on the correlation length  of dark matter
   and, hence, on the  the expansion of the universe.
This theory predicts that the size of the  galaxies increases as
the Hubble radius of the universe even without merging, which agrees well with the recent observational data.
\end{abstract}

 \maketitle
Dark matter (DM)
and dark energy~\cite{DMreview} are two of most important
unsolved puzzles in modern physics and cosmology.
Identification of one DM particle
species by a direct detection experiment such as LHC or
DAMA~\cite{dama}
 is not enough to fully solve the dark matter problem,
because there can be multiple species of DM and
 we have to explain the observed  distribution of DM  in the universe.
  Although the
cold dark matter (CDM) with the cosmological constant (i.e.,
$\Lambda$CDM) model   is popular and remarkably successful in
explaining the large scale structure of the universe, it seems to
encounter problems on the scale of  galactic or sub-galactic
structures~\cite{DMmodels}. Numerical simulations with
$\Lambda$CDM model usually predict a cusped central halo DM
density and too many satellite galaxies compared to  astronomical
observations~\cite{Salucci:2002nc,navarro-1996-462,deblok-2002,crisis}.

 On the other hand, it is
known  that the $cold$ dark matter model based on Bose Einstein
condensate (BEC) ~\cite{sin1}
  or scalar field dark matter (SFDM)~\cite{myhalo}
  can alleviate these problems ~\cite{corePeebles,PhysRevD.62.103517,0264-9381-17-13-101,PhysRevD.63.063506}
   and well explain the observed rotation curves of galaxies ~\cite{PhysRevD.68.023511,Boehmer:2007um}.
    Henceforth, I will designate the two models
 as the BEC/SFDM model ~\cite{guzman:024023}.
  In this model the galactic halos are like gigantic atoms where cold
boson DM particles are condensated in a single macroscopic wave
function $\psi(r)$. ( The idea of giant atoms as hypothetical
$stars$ goes back to Kaup and Ruffini~\cite{PhysRev.187.1767} and  developed by Schunck and others ( See ~\cite{Schunck:2003kk}
for a review. ).)  Similar halo DM ideas were  suggested  by many authors
~\cite{Schunck:1998nq,Balakrishna:1998pa,PhysRevLett.84.3037,Mielke:2006iw,PhysRevD.64.123528,PhysRevD.65.083514,
repulsive,Fuzzy,corePeebles,Nontopological,PhysRevD.62.103517,Alcubierre:2001ea,
Fuchs:2004xe,Matos:2001ps,0264-9381-18-17-101,PhysRevD.63.125016,Julien,moffat-2006,Sikivie:2009qn,mielkeaxion}.
(See ~\cite{Schunck:2003kk,myreview} and references therein.) This BEC/SFDM
model, now often known  as the fuzzy DM model, is a variant of the CDM
model. It is more about the state of DM particles rather than the DM
particle itself.

The recent observations~\cite{daddi,compact,evolution} of the size evolution of massive galaxies even deepen
mysteries of DM and formation of galaxies.
In ~\cite{evolution}, using the combined capabilities of earth-bound telescopes and the Hubble space telescope,
the size evolution of 831 very massive galaxies since $z\sim 2$ is investigated.
In ~\cite{compact} it is observed that massive quiescent galaxies at $z\sim 3$ have a very small  median
effective radius $r=0.9 ~kpc$.
According to the observations, very massive ($\ge 10^{11}M_\odot$)
 galaxies  have a factor of  about $5$
 smaller size in the past ($z\sim 2$)
than their counterparts today. These compact but massive galaxies are puzzling in the
context of $\Lambda$CDM model, because the size evolution is usually
attributed  to a hierarchical merging process of small galaxies to form a
larger galaxy. Thus, it is  expected that small early   galaxies
have small mass too. The finding of
 compact but very  massive early galaxies which have disappeared
   calls this
interpretation in question, since these galaxies had almost
reached the maximum mass limit  observed today and might not have
been experienced significant merging after $z\sim2$~\cite{daddi}.
Various mechanisms depending on visible matter could  change the
size-mass relation by a factor of $\sim 2$ but not a factor of
$\sim 5$~\cite{compact}.

In this paper, it is  suggested that this galaxy size evolution
problem can be also solved in the BEC/SFDM model, if the
correlation length of the DM condensate is time dependent. First,
let me briefly review the BEC/SFDM model. In 1992, to explain the
observed galactic rotation curves, Sin \cite{sin1,sin2} suggested
that galactic halos are   astronomical objects in BEC  of ultra
light (mass $m\simeq 10^{-24}~eV$) DM particles such as pseudo
Nambu-Goldstone boson (PNGB). In this model  the halo cold boson DM particles are condensated in a
single macroscopic wave function $\psi(r)$
 and the quantum mechanical uncertainty principle
prevents the halos from
 self-gravitational collapse, while in usual CDM models DM particles move independently and incoherently.
 $\psi(r)$ satisfies
the non-linear Schr\"{o}dinger equation
with the Newtonian gravity;

\beq
i\hbar\partial_t \psi =E\psi =
-\frac{\hbar^2}{2m}\nabla^2\psi+
 V \psi(r),
\label{sch}
\eeq
where $E$ is the energy of each DM particle.
The gravitational potential is given by
\beq
V(r)= \int^{r}_0 dr'\frac{Gm}{r'^2}\int^{r'}_0 dr'' 4\pi r''^2
(\rho_{vis}(r'')+ \rho_{DM}(r'') ) +V_0,
\label{V}
\eeq
where DM density
$\rho_{DM}(r)=M_0 |\psi(r)|^2$
and $\rho_{vis}$ is the visible matter (i.e., stars and gas) density.
$M_0$ is a  mass parameter and $V_0$ is a constant.
According to the model, the condensation of DM particles
 of which huge Compton wavelength
$\lambda_{c} = {2\pi \hbar}/{mc} \sim 10~pc$, i.e. $m \simeq
10^{-24} eV$, is responsible for the halo formation.

 The author and Koh ~\cite{kps,myhalo} generalized Sin's BEC model
 in the context of quantum field theory and the general relativity and
 suggested that DM can be described as a coherent scalar field (i.e., SFDM).
In this model the  BEC DM halos are giant boson stars (boson halos~\cite{Schunck:2003kk,review,review2}) described by
 a  complex scalar field $\phi$ having a typical action
\beq
\label{action}
 S=\int \sqrt{-g} d^4x[\frac{-R}{16\pi G}
-\frac{g^{\mu\nu}} {2} \phi^*_{;\mu}\phi_{;\nu}
 -U(\phi)]
\eeq
with a  potential
$U(\phi)=\frac{m^2}{2}|\phi|^2+\frac{\lambda}{4}|\phi|^4$.
In this paper, the case with $\lambda=0$ will be considered for simplicity.
The spherical symmetric metric is $ds^2=-e^{\nu(r)} dt^2+e^{\lambda(r)} dr^2+r^2d\Omega^2$,
where $r$ is the radial coordinate.

In the BEC/SFDM theory, due to the uncertainty principle
$\Delta x \Delta p \ge \hbar/2$, the characteristic length scale  of a BEC DM halo $\xi$ is inversely proportional
 to the mass of DM particles, i.e.,  $\xi\sim \Delta x \sim \hbar / \Delta p\sim \hbar /m \Delta v\sim 10^{-3}c~ \hbar/m$,
   where $\Delta v$ is the
 velocity dispersion of DM particles and $c$ is the light velocity.
This
$\xi$ should be comparable with the characteristic length scale
of the DM condensate. (For DM halos or boson stars, the correlation length or de Broglie wavelength  of the condensate
is more suitable for the length scale than the Compton wavelength $\lambda_c\sim 1/m$ ~\cite{myhalo,Silverman:May}.)
From this one can also obtain the minimum mass for galaxies
$M_c=\frac{\hbar^2}{G\xi m^2}\simeq 10^{7}M_\odot$,
which is recently observed~\cite{Lee:minmass}.

 The temperature-dependent correlation length of DM
 is about the thermal de Broglie length $\xi(T)\simeq 2\pi \hbar~c /kT$
of the DM condensate~\cite{Silverman:May}, which is an increasing
function of the time.  This leads to a surprising possibility
that the sizes of a DM halo and  a galaxy embedded in the halo are  slowly increasing functions of the time
even without merging. The DM halo provides a potential well to
trap the visible galactic matter such as stars and gas.
The extension of the halo induces the extension of a visible part of
the galaxy. To see this
consider the  Newtonian limit of the equations of motions from the action in  Eq. (\ref{action}), or,
  Eq. (\ref{sch}), which
 can be  written in the dimensionless form as
 \beq
 \left\{ \begin{array}{l}
 \nabla _{} ^2 {V} = (\sigma ^2+{\rho}_{vis})  \\
  \nabla ^2 \sigma  = 2{(V-E)}\sigma    \\
\end{array} \right. .  \\
\label{all}
 \eeq
Here  $\sigma=\sqrt{4\pi G}e^{-iEt}\psi/c^2$ is
a dimensionless form of the wave function $\psi$ (or the scalar field $\phi$)
and ${r}\rightarrow mcr/\hbar$,  ${t}\rightarrow mc^2 t/\hbar$~\cite{myhalo} and
all other quantities are dimensionless too
from now on.
Since galaxies are dominated by DM,
we will ignore $\rho_{vis}$ and assume that visible matter passively moves inside the potential well $V$
of the DM halo.
The equations above have  approximate solutions~\cite{Supermassive}
 \beq
 \left\{ \begin{array}{l}
 \sigma(r)\simeq \sigma(0)\left(1+\frac{(V(0)-E) r^2}{6}\right)    \\
 V(r)\simeq V(0)+\frac{\sigma(0)^2 r^2}{6}     \\
\end{array} \right. .  \\
\label{all2}
 \eeq

To see the size evolution for the most massive galaxies,
 we need to calculate the DM distribution in their halos
for a fixed galaxy mass
$M\equiv \int_0^{\infty}
 4 \pi r^2 (\sigma(r)^2 + \rho_{vis}(r)) dr\simeq O(|\sigma|^2 r^3)$.
Thus, for a constant $M$ and under the scaling $\xi\rightarrow l
\xi$ corresponding to the increase of the correlation length, the
other parameters for DM halos scale as $r\rightarrow l r$, $\sigma
\rightarrow l^{-3/2} \sigma$ and $V\sim M/r\rightarrow V/l$. This
means, under this scaling, the gravitational potential well
becomes shallower  and the size of DM halos $r_{DM}$ increases
like $l$. We can also assume that the total energy (kinetic +
potential) of a star $E_{st}$  measured from $V(0)$  is conserved
during the extension. The star orbits
 around the galactic center within the gravitational  potential of the halo
$V$. Note that $E_{st}$ is different from the energy of a DM
particle, $E$. The visible radius of a galaxy  can be defined by an
orbital radius of outermost stars  $r_{*}$, which can be defined as
the position satisfying the condition $E_{st}=V(r=r_{*})$, i.e.,
the position where the kinetic energy of the stars is zero. From Eq.
(\ref{all2}) and the energy conservation one can  see that
$E_{st}=V(r_*)=\sigma (0)^2 r_*^2/6$ and $r_*=\sqrt{6
E_{st}}/\sigma(0)$. Since
 $\sigma (0) \rightarrow l^{-3/2} \sigma(0)$, the orbital radius  of the star scales as $r_{*}\rightarrow
l^{3/2} r_{*}$.
 It means that the visible radius of galaxy
follows approximately  the $3/2$ power of the size of its dark
matter halo. (The extension of the halo also reduces the density of matter
and the rotation velocity of stars. This
is also observed~\cite{evolution}. Since the observed `extension
speed' of the galaxy is very low
($O(kpc/10^{10}~yrs)$)~\cite{evolution}, we can treat the
expansion  as an adiabatic one. Thus, the temperature is inversely
proportional to the scale factor $R$ of the universe, i.e., $T\sim
1/R$, as usual.)

\begin{figure}[htbp]
\includegraphics[width=0.4\textwidth]{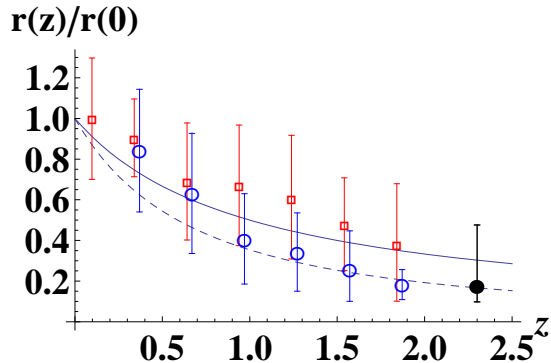}
\caption{ (Color online)
Observed size evolution $r(z)$ of the massive galaxies versus the redshift $z$.
The red squares denotes the size evolution of spheroid-like galaxies
and the blue circles are for disc-like galaxies (Data from Fig. 9. of Ref. \cite{evolution} ).
The black dot represents the typical compact galaxy at $z=2.3$ in Ref. \cite{compact}.
The line represents our theoretical prediction  $r_{DM}(z)/r_{DM}(0)=1/(1+z)$ for DM halos.
The dashed line represents the predicted
size evolution of the $visible$ parts of the galaxies  $r_{*}(z)/r_{*}(0)=1/(1+z)^{3/2}$
which agrees well with the observational data for the disc-like galaxies.}
 \label{r}
\end{figure}

Collecting all together, we obtain a simple relation between the size parameter $r_{DM}(z)$ of DM halos
and the redshift $z$;
\beq
\label{rz}
\frac{r_{DM}(z)}{r_{DM}(0)}=\frac{\xi(z)}{\xi(0)}=\frac{T(0)}{T(z)}=\frac{R(z)}{R(0)}=\frac{1}{1+z}.
\eeq
Here $r_{DM}(z)$ denotes the effective radius of a DM halo at $z$, and $r_{DM}(0)$
denotes that  at present.
Thus, the size of $visible$ galaxies $r_{*}$ evolves as
\beq
\label{rz2}
\frac{r_{*}(z)}{r_{*}(0)}=\left(\frac{r_{DM}(z)}{r_{DM}(0)}\right)^{3/2}=\left(\frac{R(z)}{R(0)}\right)^{\frac{3}{2}}=
\left(\frac{1}{1+z}\right)^{\frac{3}{2}}.
\eeq
Since the Hubble radius $H^{-1}(t)\sim t$ and $R(t)\sim t^{\frac{2}{3}}$ during the matter dominated era,
the size of the most massive galaxies increases as the Hubble parameter at the first order.
Fig. 1 shows
the observed size evolution $r(z)$ of the massive galaxies~\cite{evolution,compact} versus the redshift $z$.
The theoretical prediction ($r_{*}(z)$ in Eq. (\ref{rz2}))
is well coincident with the observational data.
The black dot represents the typical parameter for 9 galaxies in Ref. \cite{compact} for which
I used the typical values $z=2.3, r(z)=0.9 ~kpc,$ and $ r(0)=5~kpc$ from the paper.
For the error bar for the dot,
 I used the effective radiuses of the smallest and the largest galaxies in the table 1 of
Ref. ~\cite{compact}.
Although there are still many theoretical and observational uncertainties, the coincidence
between the theoretical prediction and
the data for disk-like galaxies is remarkable, especially for high $z$.
The  small discrepancy  can be attributed to the systematic observational uncertainties
and ignoring of possible merging history.
It is unclear why disc-like galaxies and spheroid-like galaxies
seem to show rather different size evolutions. We need more observations
to determine whether the difference is real or not.

 To be more illustrative, we perform a numerical study using the shooting method~\cite{myhalo}.
 Fig. 2 shows the result of our numerical study with boundary conditions
$dV/dr(0)=0,V(\infty)=0 $ and $d\sigma/dr(0)=0$. We consider 3 cases with
the  parameters
$(\sigma(0) = 5\times 10^{-7}, V(0)= -3.678\times 10^{-7},E=-1.52\times 10^{-8}),
(\sigma(0) = 3.21\times 10^{-7}, V(0)= -2.72\times 10^{-7},E=-5.71\times 10^{-8})$, and $
(\sigma(0) = 2.17\times 10^{-7}, V(0)= -2.04\times 10^{-7},E=-6.23\times 10^{-8})$, respectively, from
the top to the bottom for $\sigma(0)$.
With $E_{st}=10^{-7}$ we changed the length scale as $l=1,1.5,2$ for 3 cases
 and obtained  $r_*=1736,2861,4590$,respectively.
The  masses  $M=0.013,0.0125,0.0124$ are similar for all 3 cases.
The numerical  results support the theoretical argument above.

We need to check the reliability of the Newtonian approximation used to derive the Eqs. (\ref{rz})
and (\ref{rz2}).
From the action in Eq. (\ref{action}) and by defining  $\sigma\equiv \sqrt{4\pi G} e^{-i\omega t}\phi$ one can obtain
the dimensionless versions of the scalar field equation and the Einstein equation~\cite{colpi},
which can be reduced to the equation in Eq. (\ref{sch})
in the Newtonian limit~\cite{myhalo}.
Since the typical compact galaxies observed have mass $M\sim 10^{11}M_\odot$ and size $R\sim 1 kpc$,
the dimensionless gravitational potential $V\sim M/R\sim 4.78\times 10^{-6} $
and the expected rotation velocity dispersion is  $v_{rot}=O(\sqrt{V})\simeq 0.0022 c\simeq 650 km/s$,
which is comparable with the recent observational data $v_{rot}\simeq 510^{+165}_{-90} km/s$ by Dokkum et al~\cite{vanDokkum:2009rg}.
Thus, the compact early galaxies are basically non-relativistic objects
and the Newtonian approximation is good for these galaxies.

 Schunck, et al pointed out that the effect of pressure generated by scalar fields should
 be included  in
the rotation curves~\cite{Schunck:1998nq,Schunck-1997}.
The rotation velocity given by the circular geodesics is
$v_{rot}=r d\nu/dr e^\nu /2\simeq M(r)/r + p_r r^2 e^{\lambda+\nu}/2 $,
where the second term denotes the contribution of the radial pressure $p_r=\rho-U
=\frac{1}{2}(\omega^2 \sigma^2 e^{-\nu}+\sigma'^2 e^{-\lambda}-U)$ of
the DM field. Since $\sigma\simeq O(10^{-7}) \ll 1$ and $v_{rot}\ll c$
we can use the weak field approximation in Ref~\cite{free,PhysRevD.52.5724}.
A relevant parameter for this approximation is $\epsilon \equiv (1-\omega^2/m^2)^{1/2}\ll 1$.
In this limit $\omega\simeq m$, $d/dr\sim \epsilon$ and $e^{\lambda,\nu}\rightarrow 1$.
Thus, we obtain (dimensionless) $p_r\simeq \frac{1}{2}( \sigma^2 +\epsilon^2 \sigma^2 -\sigma^2)=O(\epsilon^2 \sigma^2)\ll \rho\sim O(\sigma^2)$
and neglecting the contribution from $p_r$ in our model (with $U=m^2 \phi^2/2$) is a good approximation
for these galaxies.
This is different from the case of the model with $massless$ scalar DM particles, where $U=0$~\cite{Schunck-1997,Schunck:1998nq}.

\begin{figure}[htbp]
\includegraphics[width=0.4\textwidth]{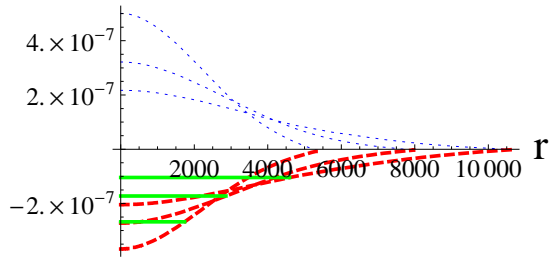}
\caption{ (Color online)
The dark matter field $\sigma$ (blue dotted lines),
the energy level of visible matter (green thick lines)   and
the gravitational potential $V$ (red dashed lines)
for a galaxy as a function of
distance  $r$ from the halo center.
As the universe expands, the temperature of DM decreases and its correlation length $\xi$
increases. This induces the increase of the size of the visible galaxy ($r_*$) represented by the green lines. }
 \label{r}
\end{figure}

We have shown analytically and numerically that the BEC/SFDM theory could explain the  observed
size evolution of the most massive galaxies.
In our theory the size of galaxies can increase
not only by merging or accretion of small galaxies but also by the increase of the length
scale of  DM halos itself.
For small and medium sized galaxies, it is hard to distinguish these two effects.
This may explain why the pure size extension without merging
is observed only recently and only for the most massive galaxies.
Our theory explains how these massive galaxies were so dense in the past
and reached the size of the massive galaxies  today.

 A self-gravitational redshift effect of the DM halo may contribute to the relation in Eq. (\ref{rz}) or
Eq. (\ref{rz2}).
The observed redshift of the galaxies could be a
combination of the cosmological ($1+z$) and the gravitational redshift ($1+z_g$), i.e., $(1 + z)(1 + z_g)$ ~\cite{Schunck:1998nq}.
The gravitational redshift parameter for the boson halo is given by ~\cite{Schunck:1998nq}
$z_g=e^{(\nu(\infty)-\nu (r))/2}-1\simeq V$ in the weak gravitation limit.
Since $V\simeq O(10^{-6})$, the correction from $z_g$ is negligible for galactic DM halos in this model ~\cite{Mielke:2006iw}
compared to the cosmological redshift.
 A time-varying gravitational constant  or the gravitational memory in Brans-
Dicke models could mimic the extension of the galaxies~\cite{memory,refree}.

In conclusion, the idea that
DM is in BEC seems to provide us a new way to explain
 not only  the CDM problems
but also  the galaxy evolution.
From this perspective it is important to determine
the exact size evolution of the galaxies by future observation.
\\

\section*{ ACKNOWLEDGMENTS }
This work was supported in part by the topical research
program (2009 -T-1) of Asia Pacific Center for
Theoretical Physics.

\vskip 5.4mm


\end{document}